\documentclass[12pt,a4paper]{article}
  \usepackage{amssymb}
  \usepackage{graphicx}
   \usepackage{latexsym}
    \usepackage{mathrsfs}
     \usepackage{feynmf}
       \usepackage{balance}
    \usepackage{amsfonts}
      \usepackage{color}
      \usepackage{geometry}
\title{Localized waves in plasmas at varying magnetic field }
\author{H. R. Pakzad\thanks{Email: ttaranomm83@yahoo.com} ,
Parvin Eslami \thanks{Email: eslami@ferdowsi.um.ac.ir} \\ and Kurosh Javidan \thanks{Email: Javidan@um.ac.ir} \\\\
\small{Department of Physics, Ferdowsi University of Mashhad}\\
\small{91775-1436  Mashhad, Iran} }
\date{}
\begin{document}
\maketitle
\begin{abstract}
By considering the continuity, Navier-Stoks and Poisson's equations in a non-relativistic frame work for plasmas, we study the behavior of small amplitude ion acoustic solitary waves in plasmas under the influence of a varying magnetic field. The result is a nonlinear wave equation which complies with the KdV-Burgers (KdVB) equation, surprisingly in the absence of thermal pressure or any dissipative effects. We show that the complete set of equations, by considering the varying magnetic field, create solitary waves which radiate energy during their travelling in the medium. An interesting result is the existence of small amplitude localized shock profiles beside the solitary waves. Properties of this solitaire solution is studied by considering different values for the environmental characters.
\end{abstract}

\section*{I. Introduction}
Soon after establishing the idea of the existence of localized waves in plasmas \cite{1, 2, 3}, study of nonlinear waves in such media has become one of the most important topics in plasma physics. During this times, a huge number of researches have been presented. Authors have investigated properties of linear and nonlinear localized waves in coupled magnetized and unmagnetized plasmas containing variety of components with different distribution functions \cite{P1, S1, C1, J1} . Meanwhile there has been reported several experimental tests and observations proving the existence of solitary waves and shock profiles in laboratory and astrophysical (and atmospheric) plasmas \cite{4, 5, 6, 7, 8, 9}. However most of situations with different conditions have been investigated, but still several papers are publishing in related journals containing newer features.\\

There are several mathematical methods for investigating plasma environments. Numerical simulations based on Magnetohydrodynamic equations of motion for plasma constituents provides the most accurate outcomes. But this method is not able to present good functional knowledge. Thus semi analytical methods like Sagdeev pseudo potential and/or reductive perturbation technique are widely used by researchers. Indeed most of published papers have used these two semi analytical methods. Results can be completed by considering stability conditions which shows permissible values for plasma parameters.\

Surprisingly, there are only a very few studies on the collective behavior and long range interactions in plasmas with non-uniform parameters \cite{10, 11, 12} especially spatial varying magnetic field, while in realistic situations, magnetic field is not constant at all. Presented works have tried to treat the problem using numerical calculations. It may be because of some difficulties in using Sagdeev potential and reductive perturbation method in such situations. Motivated by this problem, it seems to be important to examine semi analytical methods to find compact approximate solution for a plasma in a non-uniform magnetic field.   

Outlines of this paper are as follows: in the next section we briefly present the magnetohydrodynamic equations of non-relativistic plasmas in non-uniform magnetic field. In the section III, evolution equation of small amplitude waves in electron-ion plasma at non-uniform magnetic field will be derived, using the reductive perturbation method. Localized solution of derived equation is discussed and time evolution of the solution in the medium is analysed in this section too. The last section is devoted to some concluding remarks.

\section*{II. Magnetohydrodynamic equations of plasmas in space dependent magnetic field }

Plasma species have different charges and due to the magnetic field, they find different trajectories in the media. Indeed, the whole fluid motion is supported  by  a certain velocity, every particle can take a specific velocity which are different and the effect of this difference is important as a source of perturbation. This means that, we have to apply the multi fluid approach to describe the system equation of motion \cite{RA,Mag,Ma}. Separation of plasma constituents in term of their charges due to strong magnetic field, also has been proven experimentally \cite{B, LEA}.    

For describing the magnetohydrodynamical evolution, we can write a system of equations which are governed by a conservation law, the energy-momentum equations of motion and evolution of the electromagnetic field through the  Maxwell's equations\cite{3,RA,Abr, Chen}.  So the set of equations can be written as following:

 \begin{equation}\label{1}
 \frac{\partial n_{j}}{\partial t}+{\bf {\nabla}} . (n_{j} {{\bf{u}}_j})=0
 \end{equation}
\begin{equation}\label{2}
n_{j}\left(\frac{\partial }{\partial t}+{{\bf{u}}_j} . {\bf{\nabla}}\right){{\bf{u}}_j}=  -{\bf{\nabla}} p_i +n_{j}(
{\bf{E}}+{{\bf{u}}_j}\times {\bf{B}})
\end{equation}
\begin{equation}\label{5}
\frac{\partial {\bf{B}}}{\partial t}=-{\bf{\nabla}}\times{\bf{E}}
\end{equation}
\begin{equation}\label{6}
{\bf{\nabla}}\times{\bf{B}} -\epsilon\mu\frac{\partial {\bf{E}}}{\partial t}=\mu\sum\limits_{i}  n_{j}{\bf{u}}_j 
\end{equation}
\begin{equation}\label{7}
{\bf{\nabla}}.{\bf{B}}=0
\end{equation}
\begin{equation}\label{8}
{\bf{\nabla}}.{\bf{E}}=\frac{1}{\epsilon}\sum \limits_{j} n_{j}
\end{equation}
in which  $j$ runs over plasma constituents (here: $j=e,i$ for electrons and ions) Equations (\ref{1}) and (\ref{2}) are the number density continuity equation and non relativistic equation of motion, where ${ \bf{u}}_j $ denotes the velocity of electrons and ions.  Also, the equations (\ref{5})-(\ref{8}) are the Maxwell's equations with the magnetic field $ {\bf{B}} $ and the electric vector field $ {\bf{E}} $, while $\epsilon$ and $\mu $ are effective dielectric constant and magnetic permeability respectively. We can combine sets of momentum equations for plasma particles. So that, by ignoring the displacement current\cite{Ka}, the equation (\ref{6}) becomes:

\begin{equation}\label{6a}
{\bf{\nabla}}\times{\bf{B}}=\mu (n_{i}{\bf{u}}_{i} -n_{e} {\bf{u}}_{e})
\end{equation}
For solving above equations we still need another excess equation, namely the equation of state, which provides a relation between plasma pressure and other dynamical variables of the system, like the number density. There is no general analytical solution except that we solve the equations in some special cases where we have more information about the plasma system. \

In order to focus on important points of the problem, let us consider the simplest plasma, which consists of ions and thermal electrons in the presence of a space dependent slowly varying external magnetic field along the $z$ axis. As electrons are in their thermal equilibrium, evolution equations for ions are: 
 \begin{equation}\label{con}
 \frac{\partial n_i}{\partial t}+{\bf {\nabla}} . (n_i {{\bf{u_i}}})=0
 \end{equation}
\begin{equation}\label{Nav}
\frac{\partial {\bf{u_i}} }{\partial t}+{{\bf{u_i}}} . {\bf{\nabla}}{{\bf{u_i}}}= -\frac{e}{m_i}{\bf{\nabla}}\Phi+\frac{e}{m_i}{{\bf{u_i}}}\times {\bf{B}}
\end{equation}
\begin{equation}\label{Lap}
\nabla^{2} \Phi=-\frac{e}{\epsilon}\left(n_i-n_e\right)
\end{equation}
where $n_i$, $u_i$ and $\Phi$ are the ion number density, the ion mean velocity, and the electrostatic potential, respectively. We assume that the wave is propagating in the $x- z$ plane. Thus, the basic set of equations (\ref{con})-(\ref{Lap}) can be written in a normalized form as:
\begin{equation}\label{conn}
 \frac{\partial n}{\partial t}+\frac{\partial}{\partial x}(nu_x)+\frac{\partial}{\partial z}(nu_z)=0
 \end{equation}
 \begin{equation}\label{Navx}
\frac{\partial u_x }{\partial t}+\left(u_x \frac{\partial}{\partial x}+u_z \frac{\partial}{\partial z}\right) u_x = -\frac{\partial \phi}{\partial x}+b u_{y}
\end{equation}
 \begin{equation}\label{Navy}
\frac{\partial u_y }{\partial t}+\left(u_x \frac{\partial}{\partial x}+u_z \frac{\partial}{\partial z}\right) u_y = -b u_{x}
\end{equation}
 \begin{equation}\label{Navz}
\frac{\partial u_z }{\partial t}+\left(u_x \frac{\partial}{\partial x}+u_z \frac{\partial}{\partial z}\right) u_z = -\frac{\partial \phi}{\partial z}
\end{equation}
\begin{equation}\label{lap}
\frac{\partial^2 \phi}{\partial x^2}+\frac{\partial^2 \phi}{\partial z^2}=n_e - n=e^{\phi}-n
\end{equation}
in which $n$, $u$ and $\phi$ are previously defined parameters normalized by $n_i0$ (equilibrium value of ion number density), $C_i = \sqrt{\frac{k_B T_e}{m_i}}$
and $\frac{k_B T_e}{e}$ respectively, while $k_B$ is the Boltzmann's constant and $m_i$ stands as the mass of positively charged ions. The time $t$ and the distance $r$ are normalized by the ion plasma frequency $\omega _{pi} ^{-1}=\sqrt{\frac{m_i}{n_{i0} e^2}}$ and the Debye length $\lambda_D=\frac{C_i}{\omega_{pi}}$ respectively. The key variable in our problem is the space dependent parameter $b({\bf{r}})=\frac{\omega_B}{\omega_{pi}}$ where $\omega_B=\frac{eB({\bf{r}})}{m_i}$. Note that the magnetic field in our model is a function of space. The electron distribution is assumed to be Maxwellian, so it is given as: 
\begin{equation}\label{elec}
n_e=e^{\phi}
\end{equation}
In this step, we construct a weakly nonlinear theory of the electrostatic waves with small but finite amplitude which leads to a scaling of independent variables through the stretched coordinates 
\begin{eqnarray}\label{st}
&&\xi=\epsilon^{\frac{1}{2}}\left(l_x x+l_y y+l_z z-\lambda t\right) \\
&&\tau=\epsilon^{\frac{2}{3}}t \nonumber \nonumber
\end{eqnarray}
where $\epsilon$ is a small dimensionless parameter scaling the weakness of dispersion and nonlinearity, $\lambda$ is the phase velocity (to be determined later) normalized by IA speed ($C_i$ ). Parameters $l_x, l_y$ and $l_z$ are the directional cosines of the wave vector ${\bf{k}}$ along the x, y and z axes, respectively, so that $l_x^2+l_y^2+l_z^2=1$. We also expand $n, u_x, u_y $ and $u_z$ in a power series of $\epsilon$

\begin{eqnarray}\label{15}
&&n=1 +\epsilon n^{(1)}+\epsilon^2 n^{(2)}+...\nonumber\\
&&u_{x}=\epsilon^{\frac{3}{2}} u_{x}^{(1)}+\epsilon^2 u_{x}^{(2)}+...\nonumber\\
&&u_{y}=\epsilon^{\frac{3}{2}} u_{y}^{(1)}+\epsilon^2 u_{y}^{(2)}+...\\
&&u_{z}=\epsilon^{\frac{3}{2}} u_{z}^{(1)}+\epsilon^2 u_{z}^{(2)}+...\nonumber\\
&&\phi=\epsilon\phi^{(1)}+\epsilon^2 \phi^{(2)}+...\nonumber\\ \nonumber
\end{eqnarray}

Now we use the stretched coordinates (\ref{st}) and expansions (\ref{15}) in (\ref{conn})-(\ref{lap}), and collect same terms in different powers of $\epsilon$. The final result is the KdV-Burgers (KdVB) equation yields:
\begin{equation}\label{KdVB}
\frac{\partial \phi^{(1)}}{\partial \tau} +l_z \phi^{(1)} \frac{\partial \phi^{(1)}}{\partial \xi}+\frac{1}{2} l_z \left(1-l_z^2\right) \frac{\partial}{\partial \xi} \left(\frac{1}{b^2}\frac{\partial ^2\phi^{(1)}}{\partial \xi^2}\right)+\frac{1}{2}l_z \frac{\partial^3 \phi^{(1)}}{\partial \xi^3} =0
\end{equation}
It should be noted that the important parameter $"b"$ in the above equation is related to the variable magnetic field. Now we have more realistic equation which express the evolution of small amplitude IA solitary waves. In a uniform magnetic field, evolution equation is a usual KdV equation \cite{mag1, mag2, mag3}. It means that stable solitary waves in uniform magnetic fields move without any dispersion. But our derived equation shows that in a non-uniform magnetic field (as actual situations) solitary waves moving with some dispersions, related to the rate of changes in magnetic field. We will show that some background noise which is observed in moving atmospheric solitary waves (and also other similar situations) can be created due to moving localized waves in space dependent magnetic fields.   

\section*{III. Discussion }
In this section a comprehensive study on the features of localized waves is carried out. The derived localized solution (\ref{KdVB}) helps us to describe effects of different parameters of the medium on the characteristics of propagated ion acoustic solitary waves. Let us look at the derived equation (\ref{KdVB}) with more details as follows:
\begin{equation}\label{KdVBe}
\frac{\partial \phi^{(1)}}{\partial \tau} +l_z \phi^{(1)} \frac{\partial \phi^{(1)}}{\partial \xi}+\frac{1}{2} l_z \left(1+\frac{\left(1-l_z^2\right)}{b^2}\right) \frac{\partial ^3\phi^{(1)}}{\partial \xi^3}+\frac{1}{2}l_z\left(1-l_z^2\right)\frac{\partial}{\partial \xi}\left(\frac{1}{b^2}\right) \frac{\partial^2 \phi^{(1)}}{\partial \xi^2} =0
\end{equation}
As a reminder, the general form of the KdV-Burgers equation is 
\begin{equation} \label{KdVBg}
\frac{\partial \phi^{(1)}}{\partial \tau}+\gamma \phi^{(1)} \frac{\partial \phi^{(1)}}{\partial \xi}+\beta \frac{\partial ^3\phi^{(1)}}{\partial \xi^3}+ \eta \frac{\partial^2 \phi^{(1)}}{\partial \xi^2} =0
\end{equation}
where $\gamma, \beta $ and $\eta$ are nonlinear, dispersion and dissipation coefficients. It may be noted that, the coefficient of the dissipative term in (\ref{KdVBe}) is a spatial function, through non-uniform magnetic field encoded in $b(\xi)$. In the absence of dissipative term (the KdV equation), nonlinear and dispersion effects may cancel out each others so that the the result becomes a stable solitary wave as:
\begin{equation} \label{KdVsol}
\phi^{(1)}=\phi^0 sech^2\left(\frac{\chi}{W} \right)
\end{equation}
where $\chi=\xi-u\tau$, $\phi^0=\frac{3U}{\beta}$ and $W=2\sqrt{\frac{\beta}{U}}$ while $U$ is the soliton velocity.

 But when the dissipative term exists, solitary solution changes into a shock profile, where soliton emerges amounts of energy radiation during its evolution in dissipative medium. Equation (\ref{KdVBe}) shows that the dissipative term appears when the magnetic field is a function of spatial position. Therefore we expect that stable solitudes radiate energy as shock waves while moving in a variable magnetic field. It is an important result in most of plasmas like the earth atmosphere. It is clear that the magnetic field is not constant around the earth, thus we expect solitary waves change into shock waves while moving in the atmosphere. \

In order to study effects of varying magnetic field on the behaviour of solitary solutions, we consider a simple perturbation in the magnetic field as $B=B_0 \left( 1+B_p e^{-\alpha \xi^2}\right)$. Normalized function of the magnetic field which appears in the equation of motion has the  form of $b=\alpha\left(1+\beta e^{-\delta \xi^2}\right)$. At positions far from the perturbation, the magnetic field is constant ($B=B_0$). Therefore at infinity, we have a stable solitary wave which goes toward the varying magnetic field and spoils when moving through the perturbation. There has not known exact solution for the KdVB equation with space dependent coefficients. Thus we have to use numerical calculation for simulating the evolution of the localized solution in varying magnetic field. The equation (\ref{KdVB}) has been solved using the Runge-Kutta method for time derivation and finite difference method for space derivations. The grid spacing has been chosen $\Delta \xi= 0.001$ and $0.005$ (as cross check for numerical stability of solution) and time grid spacing has been taken as $\Delta \tau=0.0001$. Figure 1 presents the solitary solution at t=0 (far from the magnetic field perturbation) and when it has been passed from the perturbation $b=0.9\left(1+0.5 e^{-0.2 x^2}\right)$. The KdV solitary wave has been located in $\xi=-20$ while the center of perturbation is $\xi=0$. Soliton has been sent toward the perturbation with speed $U=0.6$ and has passed through the varying magnetic field. Figure 1 shows the soliton when it arrives at $\xi=+12$. Interestingly, we find that an extra backward propagating shock structure is created at the location of perturbation. 
\begin{figure}[htp]\label{fig1}
\centerline{\begin{tabular}{cc}
\includegraphics[width=10 cm, height= 10 cm]{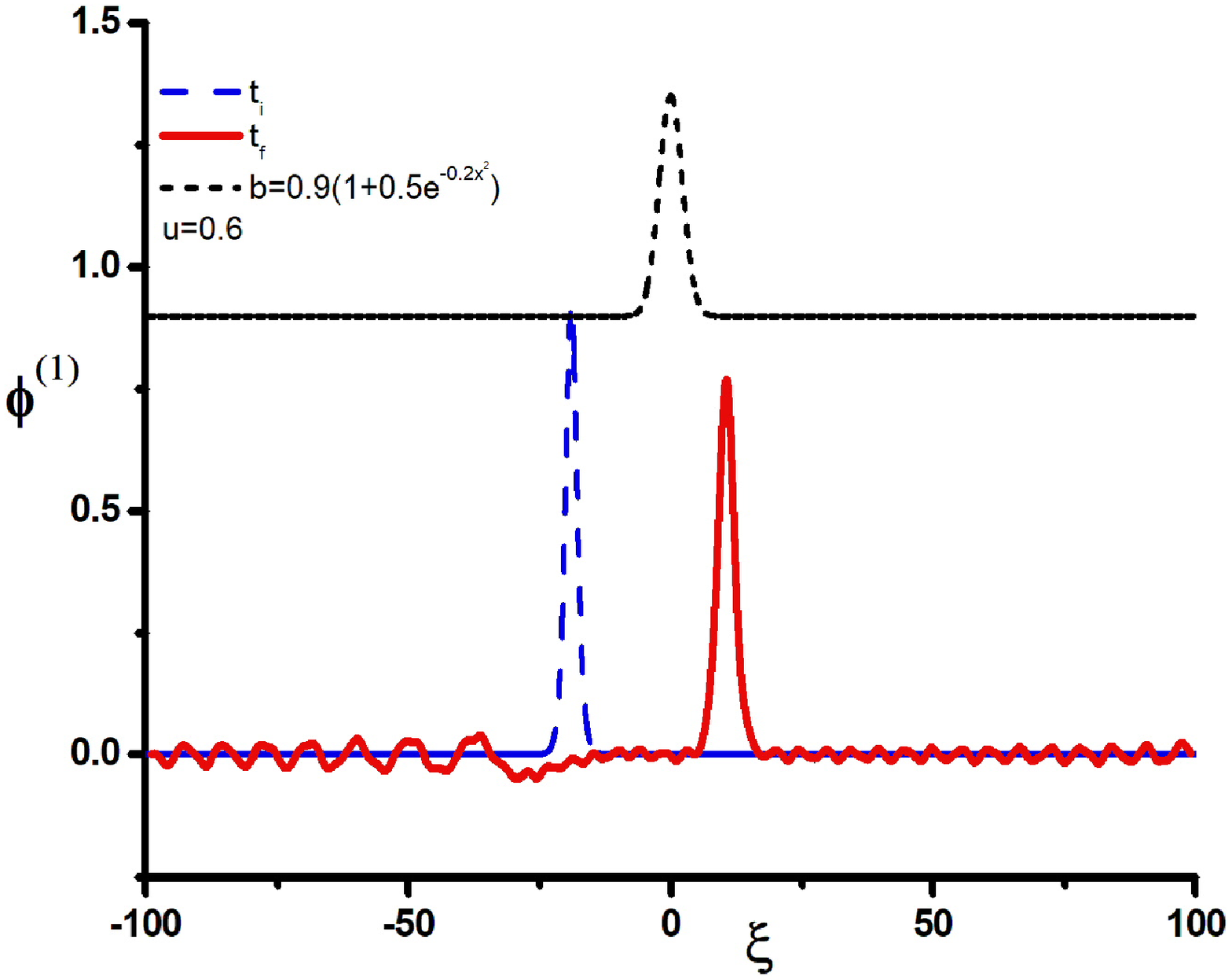}
\end{tabular}}
 \caption{\footnotesize
Time evolution of the normalized  KdV solitary solution while interacting with the magnetic field $b=0.9\left(1+0.5 e^{-0.2 x^2}\right)$.}
\end{figure}
Amount of radiated energy (and shock amplitude) becomes greater when changing rate of magnetic field perturbation increases. Figure 2 demonstrates solitary wave during the interaction with magnetic field perturbation $b=0.4\left(1+0.2 e^{-0.1 x^2}\right)$.  
\begin{figure}[htp]\label{fig2}
\centerline{\begin{tabular}{cc}
\includegraphics[width=10 cm, height= 10 cm]{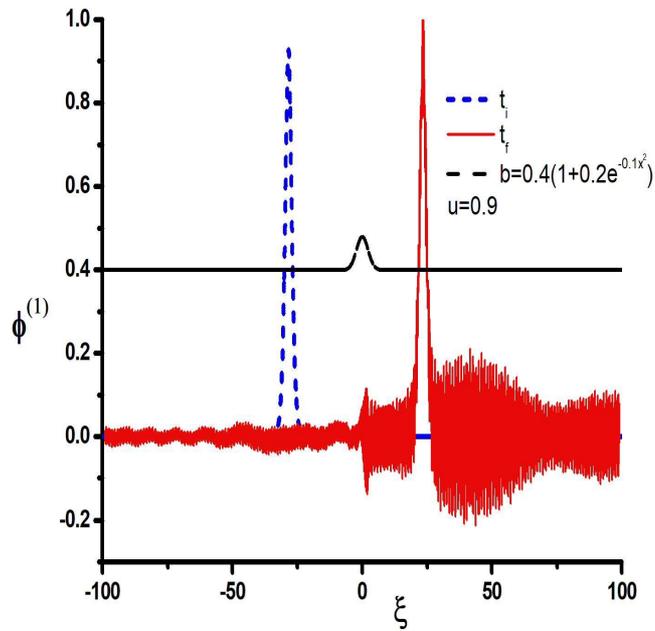}
\end{tabular}}
 \caption{\footnotesize
Time evolution of the normalized KdV solitary solution while interacting with the perturbation in magnetic field $B=0.4\left(1+0.2 e^{-0.1 x^2}\right)$.}
\end{figure}
For greater values of dissipative coefficient, shock profile becomes dominant and in such cases, solitary wave moves in a noisy medium while reducing its energy due to radiation.
We learn from the equation (\ref{KdVBe}) that, the cosine angle $l_z$ plays a serious role in the characteristics of solitary waves and their propagation. In small angles between the direction of moving soliton and the direction of magnetic field ($l_z \approx 1$) dissipative term goes toward zero and thus the dissipation effects reduces, while for greater angles (about $\cos ^{-1}\left(\frac{\sqrt{3}}{3}\right)$) the dissipative term becomes important, i.e. solitons emerge noticeable amounts of radiative energies. \

In order to find a physical sense, we present a crude estimation for the evolution of a small amplitude solitary wave in the earth atmosphere. The earth magnetic field varies  from $30000  \thinspace nT$ in  0(latitude):+60(altitude) to $45000  \thinspace nT$ in 10:+90 \cite{UKM}, where the magnetic field is almost horizontal. The electron (ion) density and temperature at height 100 Km are about $n_{e0}\approx 2 \times 10^{11} \left( n_{i0}= 10^{11} \right) m^{-3}$ \cite{Rich, G1} and $T_{e}\approx 2000-3000 ^{\circ}K \left(T_{i}=800-1200 ^{\circ}K\right)$ \cite{G2, Kohn}. Through mentioned values, the space dependent magnetic field can be estimated as $b \approx 2400\left(1+0.5 e^{-10^{-6}\xi^2}\right)$ in which distance is normalized by the Debye length (about 11 Km). As the magnetic field is so extended, initial position of the solitary wave has been located very far from the center of perturbation. An interesting point in the figure 3 is the reduction of the solitary amplitude during its travelling in the perturbation. It should be noted that, in above area of the earth atmosphere, $l_z$ varies from 0.98 to 0.87 \cite{UKM}, but we have taken the fixed value 0.9 in our calculations. This figure clearly indicates that solitary amplitude reduces while moving in a varying magnetic field, however it is able to travel long distances in the earth atmosphere. \\   
\begin{figure}[htp]\label{fig3}
\centerline{\begin{tabular}{cc}
\includegraphics[width=10 cm, height= 10 cm]{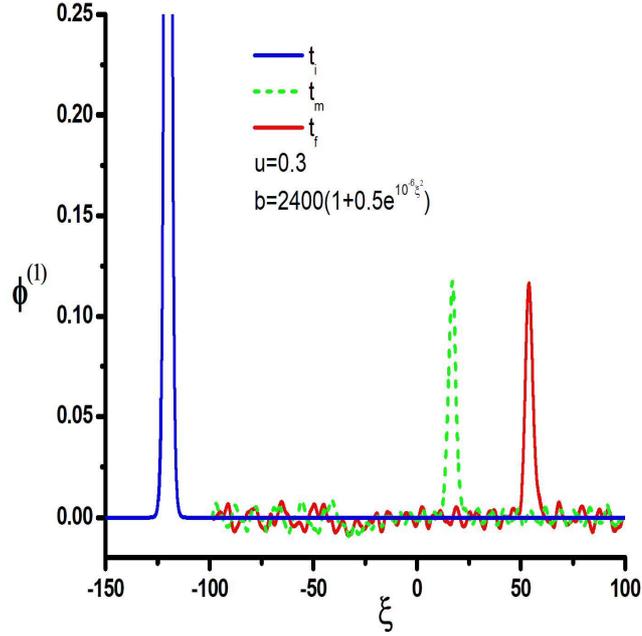}
\end{tabular}}
 \caption{\footnotesize
solitary solution travelling in the earth atmosphere, where magnetic field is the form $b \approx 2400\left(1+0.5 e^{-10^{-6}\xi^2}\right)$.}
\end{figure}

\section*{IV. Conclusions and remarks }
Propagation of small amplitude localized solution in a magnetized plasma containing electrons and ions has been studied, when the magnetic field is a varying function of space. It is a more realistic situation especially for the earth atmosphere. We have shown that a space dependent magnetic field creates new effects through adding a dissipative term to the equation of motion. Indeed the master equation for the evolution of IASW changes from the KdV equation into the KdV-Burgers type equation with a space dependent dissipation coefficient. As the equation has not known exact solution we have studied the behaviour of solitary solutions numerically.\

Our calculations show that, at the presence of varying magnetic field, solitons radiate some amount of energies during their travelling through the varying magnetic field. Radiated energy emerges as backward moving shock waves and its amplitude is very sensitive to the variation of magnetic field and also the angle between the direction of magnetic field and the soliton velocity. Moreover we have demonstrated a crude estimation for the propagation of solitary waves in the earth atmosphere. \

Our work is just first step of many investigations, which should be done. One can consider effects of dusts, hydrothermal distribution for plasma constituents, existence of warm plasma, considering multi component plasmas and many other features in astrophysical and/or laboratory plasmas. It is possible that there exist some sorts of instabilities in the propagating waves, which could be investigated. 

\section*{acknowledgement}
This work is supported by the Ferdowsi University of Mashhad under the Grant NO. $ 3/28787 $


\begin{thebibliography}{50}

\bibitem {1} O. W. Greenberg and Y. M. Trève Citation, Shock Wave and Solitary Wave Structure in a Plasma, Physics of Fluids 3, 769 (1960), doi:10.1063/1.1706124

\bibitem{2} J. G. Cordey, Solitary Waves in a CollisionFree Plasma with an Isotropic Pressure, Physics of Fluids 7, 778 (1964), doi: 10.1063/1.1711285

\bibitem{3} R. L. Smith  and N. Brice, Propagation in Multicomponent Plasmas, JOURNAL OF GEOPHYSICAL RESEARCH, VOL. 69, No. 23, (1964), DOI: 10.1029/JZ069i023p05029


\bibitem{P1} H. R. Pakzad, and K. Javidan, Obliquely propagating electron acoustic solitons in magnetized plasmas with nonextensive electrons, Nonlin. Processes Geophys., 20, 249-255, DOI: 10.5194/npg-20-249-2013, 2013. 


\bibitem{S1} S. Sultana1, I. Kourakis, N. S. Saini, and M. A. Hellberg, Oblique electrostatic excitations in a magnetized plasma in the presence of excess superthermal electrons, Physics of Plasmas 17, 032310 (2010), DOI: 10.1063/1.3322895

\bibitem{C1} P. Chatterjee, T. Saha, S. V. Muniandy, C. S. Wong, and R. Roychoudhury, Ion acoustic solitary waves and double layers in dense electron-positron-ion magnetoplasma, Physics of Plasmas 17, 012106 (2010),  DOI: 10.1063/1.3291059

\bibitem{J1}K. Javidan and D. Saadatmand, Effect of high relativistic ions on ion acoustic solitons in electron-ion-positron plasmas with nonthermal electrons and thermal positrons,  Astrophys Space Sci, (2011) 333: 471. DOI:10.1007/s10509-011-0645-6 

\bibitem{4} R. E. Ergun and et al. (18 names) FAST satellite observations of large‐amplitude solitary structures, Geophysical research letters, Volume25, Issue12
(1998) 2041-2044, DOI:10.1029/98GL00636

\bibitem{5} P. O. Dovner, A. I. Eriksson, R. Boström and B. Holback, Freja multiprobe observations of electrostatic solitary structures, Geophysical research letters, Volume Issue The Freja Project, (1994), 1827-1830, DOI: 10.1029/94GL00886 

\bibitem{6} G. O. Ludwig, J. L. Ferreira, and Y. Nakamura, Observation of Ion-Acoustic Rarefaction Solitons in a Multicomponent Plasma with Negative Ions, Phys. Rev. Lett. 52, 275 (1984) DOI: 10.1103/PhysRevLett.52.275 

\bibitem{7} Y. Nakamura and I. Tsukabayashi, Observation of Modified Korteweg—de Vries Solitons in a Multicomponent Plasma with Negative Ions, Phys. Rev. Lett. 52, 2356 (1984) 10.1103/PhysRevLett.52.2356

\bibitem{8} H. Bailung, S. K. Sharma, and Y. Nakamura, Observation of Peregrine Solitons in a Multicomponent Plasma with Negative Ions, Phys. Rev. Lett. 107, 255005 (2011) DOI: 10.1103/PhysRevLett.107.255005 

\bibitem{9} K. Stasiewicz, Theory and Observations of Slow-Mode Solitons in Space Plasmas, Phys. Rev. Lett. 93, 12  (2004) DOI: 10.1103/PhysRevLett.93.125004 

\bibitem{10} K. Javidan, P. Mohammadzadeh, Propagation of solitary waves in non uniform dusty plasmas, Astrophysics and Space Science 337(1):193-199 (2012) DOI: 10.1007/s10509-011-0800-0


\bibitem{11} G. S. Xu, V. Naulin, W. Fundamenski, J. Juul Rasmussen, A. H. Nielsen, and B. N. Wan, Intermittent convective transport carried by propagating electromagnetic filamentary structures in nonuniformly magnetized plasma, Physics of Plasmas 17, 022501 (2010), DOI:10.1063/1.3302535

\bibitem{12} J. M. Dewhurst, B. Hnat, and R. O. Dendy, The effects of nonuniform magnetic field strength on density flux and test particle transport in drift wave turbulence, Physics of Plasmas 16, 072306 (2009), DOI:10.1063/1.3177382

\bibitem {RA} A. Ghaani, K. Javidan and M. Sarbishaei, Astrophys. Space Sci {\bf{358}}, (1) 20 (2015).
\bibitem {Mag} D. A. Fogaca, S. M. Sanches and F. S. Navarra, arXiv:hep-ph/1706.02991.
\bibitem {Ma} D. M. Gomez, R. Soler and J. Terradas, ApJ {\bf{832}}, 101 ( 2016). 

\bibitem{B} S. Baboola, R. Bharuthram, and M. A. Helberg, Arbitrary-amplitude rarefactive ion-acoustic double layers in warm multi-fluid plasmas, Journal of Plasmas Physics, Volume 40, Issue 1 (2009) 163-178, DOI: 10.1017/S0022377800013180

\bibitem {LEA} J. E. Leake, V. S. Lukin, M. G. Linton, and E. T. Meier, MULTI-FLUID SIMULATIONS OF CHROMOSPHERIC MAGNETIC RECONNECTION IN A WEAKLY IONIZED REACTING PLASMA, The Astrophysical Journal, (2012) Volume 760, Number 2, DOI:10.1088/0004-637X/760/2/109

\bibitem {Abr} B. Abraham-Shrauner, (1967). Propagation of hydromagnetic waves through an anisotropic plasma, Journal of Plasma Physics, (1967) 1(03), 361. DOI:10.1017/s0022377800003354 

\bibitem {Chen} F. Chen, Introduction to Plasma Physics and Controlled Fusion, Springer, 3rd ed. (2016),  ISBN-13: 978-3319223087, ISBN-10: 9783319223087 

\bibitem {Ka} T. Kakutani, H. Ono, T. Taniuti , et al. J. Phys. Soc. Japan., (1968,) {\bf{24}}, 5

\bibitem {mag1} S. Mahmood, A. Mushtaq, and H. Saleem, Ion acoustic solitary wave in homogeneous magnetized electron-positron-ion plasmas, New Journal of Physics,  (2003) Volume 5,  DOI:10.1088/1367-2630/5/1/328

\bibitem {mag2} S. Mahmood, N. Akhtar, Ion acoustic solitary waves with adiabatic ions in magnetized electron-positron-ion plasmas, The European Physical Journal D, Volume 49, Issue 2, pp 217–222 DOI: 10.1140/epjd/e2008-00165-4

\bibitem {mag3} N. Jannat, M. Ferdousi, and A. A. Mamun, Plasma Phys. Rep. (2016) 42: 678. DOI: 10.1134/S1063780X16070059    


\bibitem {UKM} British Geological Survey, http://www.geomag.bgs.ac.uk/education/earthmag.html 

\bibitem {Rich} P. G. Richards, D. Bilitza, and D. Voglozin, Ion density calculator (IDC): A new efficient model of ionospheric ion densities, RADIO SCIENCE, VOL. 45, RS5007, (2010) DOI:10.1029/2009RS004332

\bibitem {G1} M. Ghobakhloo, M.E. Zomorrodian, K. Javidan, Distribution functions of electrons in the Earth and Venus ionospheres:Effects on the propagation of localized waves, New Astronomy, Volume 62, (2018), 115-120, DOI: 10.1016/j.newast.2018.02.001

\bibitem {G2} M. Ghobakhloo, M. E. Zomorrodian, and K. Javidan, Effects of dust polarity and nonextensive electrons on the dust-ion acoustic solitons and double layers in earth atmosphere, Advances in Space Research, Volume 61, Issue 9, (2018), 2259-2266 DOI: 10.1016/j.asr.2018.02.012

\bibitem {Kohn} W. Kohnlen,  A MODEL OF THE ELECTRON AND ION TEMPERATURES IN THE IONOSPHERE, Planet. Space Sci., (1986) Vol. 34, No. 7, pp. 609-630 



\end{thebibliography}
\end{document}